\documentclass[aps,prd,preprint,groupedaddress,longbibliography]{revtex4-1}

\usepackage{bbm}
\usepackage{color}
\usepackage{graphicx}
\usepackage{amssymb,latexsym}

\usepackage[utf8]{inputenc}

\newcommand{\newn}{x}
\newcommand{\hnewn}{m}

\newcommand{\lnn}{\sigma \frac{\hnewn_1}{\hnewn_2}}

\newcommand{\Xilor}{\Xi_{\textrm{\scriptsize Lor}}}

\newcommand{\ephys}{{
     q}}
\newcommand{\ephysphys}{{
    |\ephys_{\phys}|}}

\newcommand{\esquare}{{    p^2_{\textrm{\scriptsize eff}}}}
\newcommand{\enosquare}{{    |p_{\textrm{\scriptsize eff}}|}}

\newcommand{\noQ}{{\esquare}}

\newcommand{\phys}{\textrm{\scriptsize phys}}

\newcommand{\eel}[1]{\label{#1}\end{equation}}
\newcommand{\eeal}[1]{\label{#1}\end{eqnarray}}

\newcommand{\bel}[1]{\begin{equation}\label{#1}}
\newcommand{\bea}{\begin{eqnarray}}
\newcommand{\bean}{\begin{eqnarray}\nonumber}
\newcommand{\beal}[1]{\begin{eqnarray}\label{#1}}
\newcommand{\eea}{\end{eqnarray}}

\newcommand{\nn}{\nonumber}

\newcommand{\be}{\begin{equation}}
\newcommand{\eeq}{\end{equation}}
\newcommand{\ee}{\end{equation}}
\newcommand{\beqa}{\begin{eqnarray}}
\newcommand{\eeqa}{\end{eqnarray}}
\newcommand{\beqan}{\begin{eqnarray*}}
\newcommand{\eeqan}{\end{eqnarray*}}
\newcommand{\ba}{\begin{array}}
\newcommand{\ea}{\end{array}}

{\catcode `\@=11 \global\let\AddToReset=\@addtoreset}
\AddToReset{equation}{section}

\newcounter{mnotecount}[section]

\renewcommand{\themnotecount}{\thesection.\arabic{mnotecount}}

\newcommand{\mnote}[1]
{\protect{\stepcounter{mnotecount}}$^{\mbox{\footnotesize
$
\bullet$\themnotecount}}$ \marginpar{
\raggedright\tiny\em
$\!\!\!\!\!\!\,\bullet$\themnotecount: #1} }

\newcommand{\R}{\mathbb R}
\newcommand{\N}{\mathbb N}
\newcommand{\Z}{\mathbb Z}
\newcommand{\Q}{\mathbb Q}

\newcommand{\eq}[1]{(\ref{#1})}

\newcommand{\ptc}[1]{\mnote{{\bf ptc:}#1}}

\newcommand{\beaa}{\begin{eqnarray*}}
\newcommand{\eeaa}{\end{eqnarray*}}

\newcommand{\dnotdelta}{\mathrm{d}}

\def\ben{\begin{equation}}
\def\een{\end{equation}}
\def\bena{\begin{eqnarray}}
\def\eena{\end{eqnarray}}

\def\f(#1/#2){\frac{#1}{#2}}
\def\Frac(#1/#2){\left(\frac{#1}{#2}\right)}
\def\chris(#1-#2-#3){{\mit \Gamma}^{#1}{}_{{#2}{#3}} }
\def\tilchris(#1-#2-#3){\tilde{{\mit \Gamma}}^{#1}{}_{{#2}{#3}}}
\def\hatchris(#1-#2-#3){\hat{{\mit \Gamma}}^{#1}{}_{{#2}{#3}}}

\DeclareFontFamily{OT1}{rsfs}{}
\DeclareFontShape{OT1}{rsfs}{m}{n}{ <-7> rsfs5 <7-10> rsfs7 <10-> rsfs10}{}

\begin{document}

\title{{Compact singularity-free Kerr-Newman-de Sitter instantons}\protect\thanks{Preprint UWThPh-2016-xx}}

\author{Piotr T.\ Chru\'{s}ciel$^\dagger$}
\email[]{piotr.chrusciel@univie.ac.at}
\homepage[]{http://homepage.univie.ac.at/piotr.chrusciel}
\affiliation{Erwin Schr\"odinger Institute and
Faculty of Physics,
University of Vienna}

\author{Michael H\"orzinger }
\thanks{Supported in part by  the Austrian Science Fund (FWF) under project P 23719-N16.}
\affiliation{Faculty of Physics,
University of Vienna}

\date{\today}

\begin{abstract}
Generalising arXiv:1511.08496, we construct further families of compact Einstein-Maxwell instantons associated with the Kerr-Newman metrics with a positive cosmological constant.
\end{abstract}


\maketitle

\section{Introduction}

One of the key questions in physics is how to reconcile general relativity with quantum mechanics. After years of intense work by many researchers this problem remains as elusive as ever.

One of the proposals, how to marry gravitation with the quantum, involves path integrals over Riemannian metrics~\cite{HawkingInstantons,EQG}. In this approach one changes the signature of the metric from a Lorentzian to a Riemannian one.  This is associated with a replacement of oscillating integrands by ones which, in the best of the worlds, would be exponentially decaying. One thus hopes to obtain a framework where potentially misleading mathematical manipulations would be avoided.

Once this prescription has been adopted, one can study model problems where the path integral can be approximated by stationary phase methods. Since stationary points of the action are precisely the solutions of the field equations, this leads one to consider Riemannian solutions of the otherwise Lorentzian field equations.  As a result,
Riemannian  counterparts of Lorentzian  solutions become basic objects in Euclidean Quantum Gravity. Solutions with Riemannian signature determine  transition probabilities between states~\cite{BoothMann}. They can be used to calculate probabilities of universe creations~\cite{FarhiGuthGuven}. Riemannian solutions with negative eigenvalues indicate quantum instabilities of semi-classical solutions~\cite{ColemanErice7}. From this perspective, an exhaustive knowledge of solutions of the field equations with a Riemannian signature and interesting properties becomes one of the priorities.

The above sets the general background for our study. The specific interest in the solutions presented below can be spelled-out as follows.

First, solutions with symmetries usually play a preferred role in any theory. Indeed, explicit such solutions are easier to find than solutions without symmetries. Furthermore, their properties are easier to analyse. Finally, some of their properties often turn out to be shared by many solutions. Hence the interest of the quest for solutions with symmetries. The solutions we present are invariant under a $U(1)\times U(1)$ action. This is the simplest isometry group which does not enforce space-homogeneity, and for which there is hope to be able to obtain all solutions. In fact, black-hole uniqueness theory suggests that our solutions below exhaust the space of $U(1)\times U(1)$ symmetric compact Riemannian solutions, and we note that this question goes well beyond the intended scope of this work.

Next, recall that the Kerr-Newman solutions are of key interest in view, again, of their uniqueness properties. The corresponding solutions with positive cosmological constant, discovered by Demia\'nski and Pleba\'nski~\cite{PlebanskiDemianski} and, independently, by Carter~\cite{Carterseparable}, are similarly expected to be unique under natural conditions. Properties of Riemannian counterparts of these solutions are expected to shed light on the  quantum properties of the Lorentzian solutions within the Euclidean quantisation program.

In recent work~\cite{ChHoerzinger} we constructed a family of compact Riemannian four-dimensional manifolds, parameterised by four integers  and the value of the cosmological constant $\Lambda$, obtained by complex substitutions in the Kerr-Newman de Sitter metric. The requirement of 1) smoothness and compactness of the underlying manifold, together with the condition that 2) \emph{the original coordinates form a smooth coordinate system away from the axes of rotation}, enforced a quantisation condition on the mass, effective charge and angular momentum parameters of the associated Lorentzian manifold.
The further requirement of well-defined charged spinor fields on the Riemannian manifolds led to a quantisation condition on the electric and magnetic charge of the instanton, as well as quantisation of the charge of the spinor fields.

It came as a surprise to us that one can construct a more general class of such compact instantons, which arise if one drops the requirement 2) above. In this way one obtains a larger class of such instantons, parameterised instead by three integers and a continuous parameter. This is reported on in this paper.

Indeed, in this work we follow~\cite{ChHoerzinger} to reduce the problem to solving a set of polynomial equations for a set of free parameters arising in the solutions. We show how the requirements just described lead, after a scaling, to a problem where all free parameters but one need to take integer values. One can now use numerics to find solutions of the problem at hand. Assuming that the possibly-continuous parameter was rational, we  find a unique explicit solution for all sets of integer and rational numbers considered.

We show that all quantities of interest have finite limits as one of the integers parameterising the solutions tends to infinity.   One is tempted to think of such limits as classical limits. We derive the linear relations that are satisfied by the parameters characterising the solutions in the limit. We show that the contribution to the path integrals of solutions with large ``quantum numbers'' are exponentially suppressed in the limit.

Since our solutions provide four-dimensional compact Riemannian manifolds with constant scalar curvature, in addition to their quantum relevance they can serve as initial data for classical (4+1)-dimensional vacuum general relativity with positive cosmological constant.

\section{The solutions}

The  Kerr-Newman-de Sitter (KNdS) metric  is a solution of the   Einstein-Maxwell equations,
\bena
R_{ \mu \nu}-\frac{1}{2} g_{ \mu \nu} R +\Lambda g_{\mu\nu}
=  \frac{1}{2} ( F_{\mu \rho } F _{\nu}^{\phantom{\nu}\rho } - \frac{1}{4}F^{\alpha \beta} F_{\alpha \beta} g_{\mu \nu} )\,, \nn
\label{15III15.1}
\eena
$dF=0$, $ d \star F=0$,
where $\Lambda>0$ is the cosmological constant.
In Boyer-Lindquist coordinates, after the replacements $a \rightarrow i a $, $ t \rightarrow it$ and
$e \rightarrow i e $ the metric takes the form
\bena
g&=& \frac{\Sigma}{\Delta_r} dr^2 +  \frac{\Sigma}{\Delta_\theta} d\theta^2 + \frac{\sin^2 (\theta)}{\Xi^2 \Sigma} \Delta_\theta ( a\dnotdelta t + (r^2 -a^2) \dnotdelta \varphi)^2 \nonumber \\
 &\phantom{=}&+ \frac{1}{\Xi^2 \Sigma} \Delta_r (dt - a \sin^2(\theta) \dnotdelta \varphi)^2\,,
\label{15III15.9}
\eena
where, after setting $\lambda = \Lambda/3$, we have
$
\Sigma=r^2-a^2 \cos^2(\theta)$,
$$ \Delta_r=(r^2-a^2)\left( 1- \lambda r^2 \right) -2 M  r +p^2 - e^2 \,,
$$
$
\Delta_\theta = 1 - \lambda a^2 \cos^2(\theta)$, and $\Xi=1 - \lambda a^2$.
The Maxwell potential reads
\bena
	A =  \frac{p \, \cos(\theta)}{\Sigma} \sigma_1 + \frac{e\,r}{\Sigma} \sigma_2
\,,
\label{15III15.7}
\eena
where the one-forms $\sigma_i$, $i=1,2$, are defined as
\beaa
	\sigma_1
    & = &
 \frac{1}{\Xi} \left(- a \,\dnotdelta t - (r^2-a^2) \dnotdelta \varphi \right)
\,,
\\
	\sigma_2
    & = &
    \frac{1}{\Xi} \left(-\dnotdelta t + a  \sin^2(\theta) \dnotdelta \varphi \right)
\,.
\eeaa

When studying the metric, the magnetic charge parameter $p$ and the electric charge parameter $e$ will only appear in the combination
\bel{8IX15.5}
 \esquare:= p^2 - e^2
 \,.
\ee
The notation $\esquare$ is appropriate, in that we have \emph{not} found any  solutions with $  p^2 \le e^2$ in the family considered below.

We wish to find sets of parameters, with $a\ne 0$, so that \eq{15III15.9} is a Riemannian metric on a closed manifold. This requires a  range of the variable $r$, bounded by two \emph{first-order} zeros $r_1<r_2$ of $\Delta_r$, so that (\ref{15III15.9}) is Riemannian for $\forall r \in (r_1,r_2),\;\theta \in (0,\pi)$. Changing $r$ to its negative if necessary, one is led to the following conditions on the parameters (see~\cite{ChHoerzinger} for details)
\bena
0<|a|<r_1<r_2
\,,
 \quad a^2< \lambda^{-1}
\,,
 \quad
  \Delta_r|_{(r_1,r_2)}>0
 \,.
\label{3V15.1}
\eena
Smoothness at $r=r_i$, $i=1,2$ requires that near those points $t$ defines a $2\pi \omega$-periodic coordinate, with
\bel{7V15.4}
    \omega
    := \frac{ 2 \Xi \left(r_1^2-a^2\right) }{ \Delta_{r} ' (r_1)  }
    = - \frac{ 2 \Xi \left(r_2^2-a^2\right) }{  \Delta_{r} ' (r_2)  }
\ee
(note that $\Delta_{r} ' (r_2)$ must be negative).
Smoothness at $\sin \theta=0$ is guaranteed by introducing near $r=r_i$   new $2\pi$-periodic angular coordinates $\phi_i$  defined as
\bel{5V15.1}
 \dnotdelta \varphi
   =    \dnotdelta\phi_1 +  \frac{a  }{ a^2-r_1^2  } \dnotdelta t
   =   \dnotdelta\phi_2 +  \frac{a  }{ a^2-r_2^2 } \dnotdelta t
 \,.
\ee
From now on we forget the coordinate $\varphi$ (which will turn out not be to globally defined in general), but retain the assumption  that $t$ is a smooth global periodic coordinate  away from the rotation axes. The $2\pi$-periodicity of $t$, $\phi_1$ and $\phi_2$
imposes the condition
\bel{15V15.21}
n :=   \underbrace{\frac{a \omega}{r_1^2-a^2}}_{=:\newn_1} -  \underbrace{\frac{a \omega}{r_2^2-a^2}}_{=:\newn_2}  \in \Z^{*}
 \,.
\ee
The further condition, imposed in \cite{ChHoerzinger}, that $\varphi$ is also a globally defined coordinate,  requires that $x_1$ and $x_2$ are positive integers. However, this is not necessary, and \eq{15V15.21} suffices to obtain a smooth compact manifold.

We thus need to find five real parameters $r_1,r_2,a,M,\noQ$ satisfying the constraints \eq{3V15.1} such that \eq{7V15.4}, \eq{15V15.21} and the equations
\bena
 &
 \Delta_r(r_1) = 0 = \Delta_r(r_2)  \,,  \label{15I12.1}
 &
\eena
hold. This can be rewritten as a system of four polynomial equations for five variables, with an integer parameter $n$. One thus expects a one-parameter family of solutions for each value of $n$.

We have not been able to establish directly existence of the desired solutions. Instead, the following strategy, mimicking that of~\cite{ChHoerzinger}, turned out to be successful for all \emph{rational pairs $(\newn_1,\newn_2)$} which we implemented numerically. We believe that there exists a unique solution of the problem at hand for all pairs $(\newn_1,\newn_2)\in (\R^+)^2$ satisfying \eq{15V15.21}, so that our condition of rationality of $x_1$ and $x_2$ is \emph{not} a real feature of the set of solutions. Rather it is an ansatz that allowed us to establish, using exact computer algebra, existence of solutions of the problem at hand for all rational values of parameters that we have considered.

Thus, we proceed as follows: given $(\newn_1,\newn_2)\in (\Q^+)^2$ satisfying \eq{15V15.21}, the above reduces to a system of five polynomial equations consisting of \eq{15I12.1} together with
\bena
&
 \Delta'_r(r_1)  \newn_1=  2 a \Xi= -  \Delta'_r(r_2)  \newn_2   \,, &\label{15I12.3}\\
& (r_1^2-a^2) \newn_1 =   (r_2^2-a^2) \newn_2 \,, \label{15I12.4}
\eena
for the five real parameters $r_1,r_2,a,M,\noQ$ which should satisfy the constraints \eq{3V15.1}. Note that \eq{15I12.4} implies
$$
 \newn_1>\newn_2 \ \mbox{so that $n>0$.}
$$

For each pair $(x_1,x_2)$ of explicit rational numbers, one can use {\sc Mathematica} to obtain an equivalent  hierarchic system of polynomial equations, the first one depending only on $\noQ $, the second one only on $\noQ$ and $a$, and so forth. (We have not been able to implement this strategy when, say $x_1$ is left as a free parameter, or is not a rational number.)
We have explicitly carried-out the analysis for a subset of values satisfying $\frac{1}{20} \le \newn_2 < \newn_1\le 10^7$,
namely for the following values: $\newn_2=\frac{p}{q}$, $\newn_1=n+\newn_2$,
$
 n \in \{1,2,\dots,10\}\,, \quad p,q \in \{1,2,\dots,20\}
$,
together with
$$
 n \in \{1,2,\dots,50,60, \dots , 100,200, \dots , 1000,10^4,10^5,10^6,10^7\}\,,
$$
$$\frac{p}{q} \in \big\{\frac{1}{20},\frac{1}{19},\dots,1,2,\dots,50,60, \dots , 100,200, \dots , 1000,10^4,10^5,10^6,10^7
 \big\}
 \,.
$$
Several further ranges of parameters have been used when searching for extrema of the values of physical parameters.
Our (exact) {\sc Mathematica} calculations show that for all pairs $(\newn_1,\newn_2)\in (\Q^+)^2 $ considered there exists \emph{exactly one} solution satisfying all equations and constraints. This solution has the property that $\noQ>0$ and $M>0$.

So, each pair of rational numbers as above determines a unique set $(M,a,\noQ)$. The question then arises, whether there exist preferred values of electric and magnetic charges $e$ and $p$.
Similarly to~\cite{DuffMadore}, the existence of charged Dirac fields on the Riemannian  manifold leads to further constraints on these parameters. We summarise the argument of~\cite{ChHoerzinger}, which applies to the current setting with simple modifications: We start by noting that while the Maxwell field $F$ is smooth everywhere, the Maxwell potential \eq{15III15.7} is not. Indeed, a smooth  Maxwell potential $A$ is obtained near $\cos\theta=1$ and $r=r_i$ when $A$ is gauge-transformed to
$$
  A +  \frac{ p  }{\Xi}  \dnotdelta \phi_i - {\frac{e r_i }{ \Xi (a^2-r_i^2) } \dnotdelta t}=:A_-
  \,.
$$
A smooth field $A$ is obtained near $\cos\theta=-1$  and $r=r_i$ when $A$ is gauge-transformed to
$$
  A - \frac{ p  }{\Xi}    \dnotdelta \phi_i  - {\frac{e r_i }{ \Xi (a^2-r_i^2) } \dnotdelta t}=:A_+
  \,.
$$
If a Dirac field $\psi$ carries a charge $q_0\ne 0$, the gauge transformation which transforms  $A_-$ to $A_+$ induces a transformation of the Dirac field
$$
 \psi \mapsto \exp\left(\frac{2 i q_0   p}{\hbar \Xi } \phi_i\right) \psi
   \,.
$$
Keeping in mind that $ \phi_i$ is $2\pi $-periodic, the requirement of single-valuedness of $\psi$ results in the condition
\bel{1XII15.13}
 \frac{2   q_0   p}{\hbar \Xi} =: \hnewn_1 \in \Z^*
  \,.
\ee
Note that \eq{1XII15.13} differs from the usual Dirac quantisation formula by a factor $\Xi$.

Next,  the transition from a gauge potential  which is regular near  $r=r_1$ to a gauge potential  which is regular near  $r=r_2$ requires a gauge transformation
$$
 A\mapsto A+ \left(\frac{pa+ e r_2 }{ \Xi (a^2-r_2^2) }
 - \frac{pa+e r_1 }{ \Xi (a^2-r_1^2) }
 \right)
  \dnotdelta t
  \,.
$$
The associated transformation of the spinor field $\psi$ leads to the further conditions:
\bel{1XII15.14++}
 \frac{\hnewn_1 n}{2}\in \Z^*
  \,,
\ee
\bel{1XII15.14}
 \frac{    q_0 e \omega }{\hbar \Xi} \left(\frac{  r_1 }{    r_1^2-a^2  }  - \frac{  r_2 }{    r_2^2-a^2  }
        \right)  =: \hnewn_2 \in \Z^*
  \,.
\ee
Eliminating $q_0$ between \eq{1XII15.13} and \eq{1XII15.14} imposes a quantised relation between $p$ and $e$:
\bel{1XII15.15}
 p =   \frac{     \omega }{ 2   }
  \left(\frac{  r_1 }{    r_1^2-a^2  }  - \frac{  r_2 }{    r_2^2-a^2  }
        \right)
         \times  \frac{    \hnewn_1   }{   \hnewn_2 } \times
  e=: \lnn e
  \,.
\ee

There exist various correlations between the parameters describing the solutions. In order to study those we will use parameters which carry some physical information. For this, given a set $(M,a,\noQ)$, parameterised by a pair $(\newn_1,\newn_2)$ with $\newn_2-\newn_1 \in \N^*$  and arising from a smooth compact Riemannian solution, we will associate to it a \emph{Lorentzian partner solution}
with the same values of $M$ and $a$ and with $e$ and $p$ chosen so that \eq{1XII15.15} holds.  Taking into account the inequality $\noQ= p^2 -e^2>0$
one is led to the condition
\bel{2XII15.1}
 \lnn > 1
 \,.
\ee
Given a further pair $(\hnewn_1,\hnewn_2)$ such that \eq{1XII15.14++} and \eq{2XII15.1} hold (note that the last condition can always be achieved by choosing $\hnewn_1$ large enough), we can determine $|q_0|$, $|e|$ and $|p|$ from \eq{1XII15.13}-\eq{1XII15.15}:
\bean
 |e| &=&
  \sqrt{\frac  {\noQ }{(\lnn)^2-1}}
 \,,
  \quad
 |p| = {\lnn}\sqrt{\frac  {\noQ }{(\lnn)^2-1}}
 \,,
\\
 |q_0|
  &=
  & \frac {\hbar \Xi} 2  \sqrt{\frac{\sigma^2 \hnewn_1^2-\hnewn_2^2} {\noQ }}
 \,.
\eeal{2XII15.2asdf}
In this way we are led to a  family of Lorentzian solutions parameterised by one real number and three integers $(\newn_1, n,\hnewn_1, \hnewn_2)$ subject to the constraint \eq{2XII15.1}. It holds that $\noQ < p^2\to_{(\hnewn_1/\hnewn_2) \to \infty} {\noQ}$, $e\to_{(\hnewn_1/\hnewn_2) \to \infty} 0$, and thus $\noQ < p^2+e^2 \to_{(\hnewn_1/\hnewn_2) \to \infty} {\noQ}$.

Note that  the locations $r_i$ of the horizons of the partner solution will \emph{not} coincide with the locations $r_i$ of the rotation axes of the associated Riemannian  solutions; similarly for areas, surface gravities, etc.

There is an ambiguity in the definition of total mass of the associated Lorentzian space-time. In a Hamiltonian approach this ambiguity is related to the choice of the Killing vector field for which we calculate the Hamiltonian~\cite{CJKKerrdS}.
We will use the formulae
\begin{widetext}
\bena
	M_{\phys}= \frac{M}{ \Xilor ^2} \,, \quad
 	J_{\phys}= \frac{a M}{ \Xilor ^2} \,, \quad
 	e_{\phys}= \frac{e}{ \Xilor } \,, \quad
 	p_{\phys}= \frac{p}{ \Xilor } \,, \quad
 	\ephysphys{}= \frac{\sqrt{p^2+e^2}}{ \Xilor }
  \,,
 \quad
 \Xilor := 1 + \lambda a^2
 \,.
  \nn
\label{15VI19.1}
\eena
\end{widetext}
Here the mass of the Lorentzian solution has been defined to be the value of the Hamiltonian associated with the Killing vector field $\Xilor \partial_t +3^{-1} \Lambda a \partial_\varphi$, while the total angular momentum is defined with the Hamiltonian associated with $\partial_\varphi$.

\begin{figure}[th]
{\includegraphics[scale=.5]{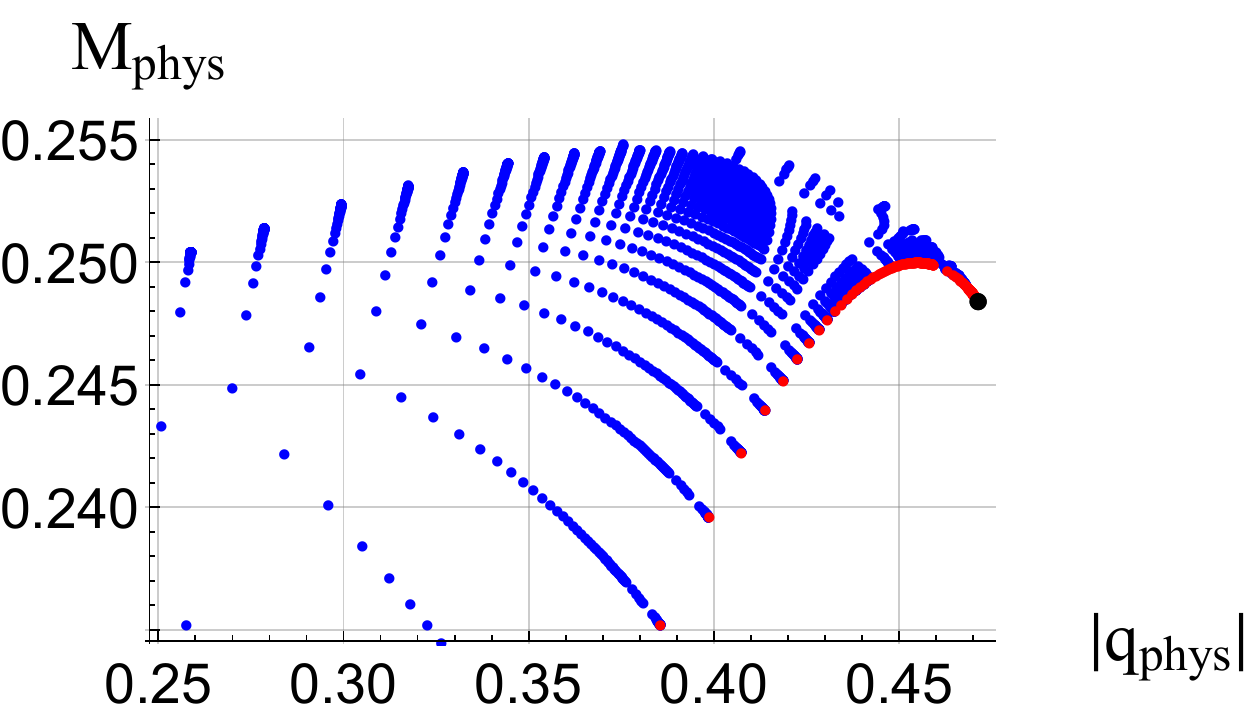}}{\includegraphics[scale=.5]{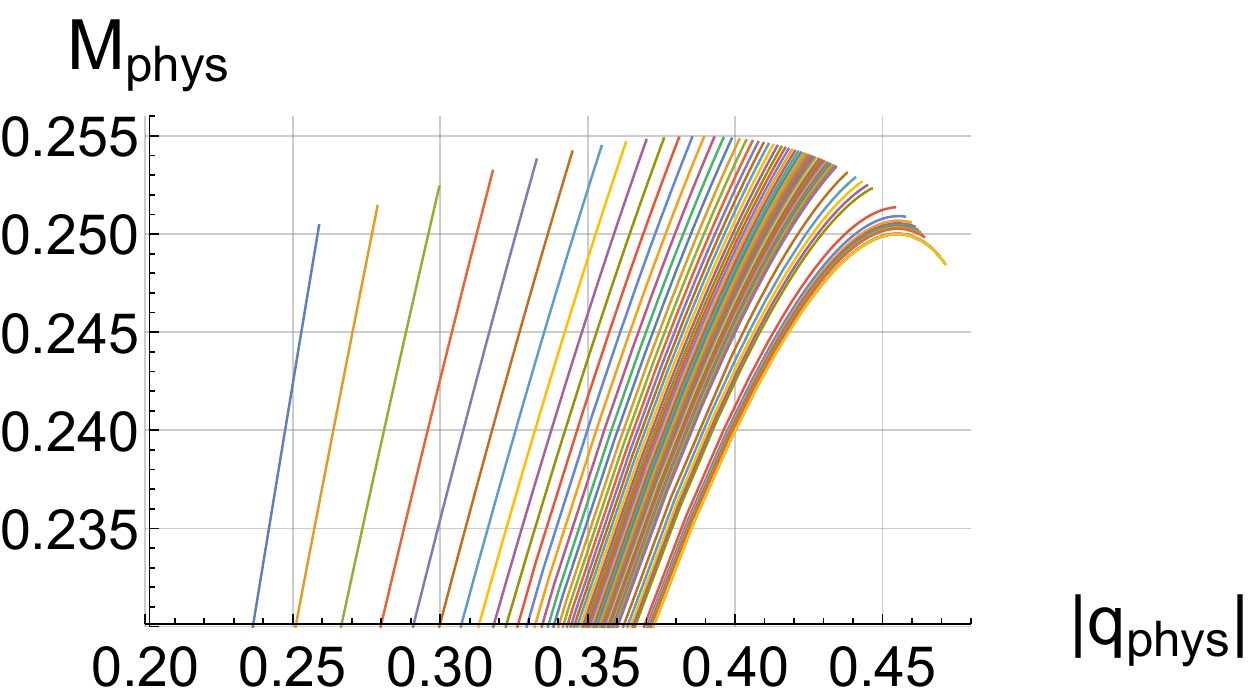}} \\
{\includegraphics[scale=.5]{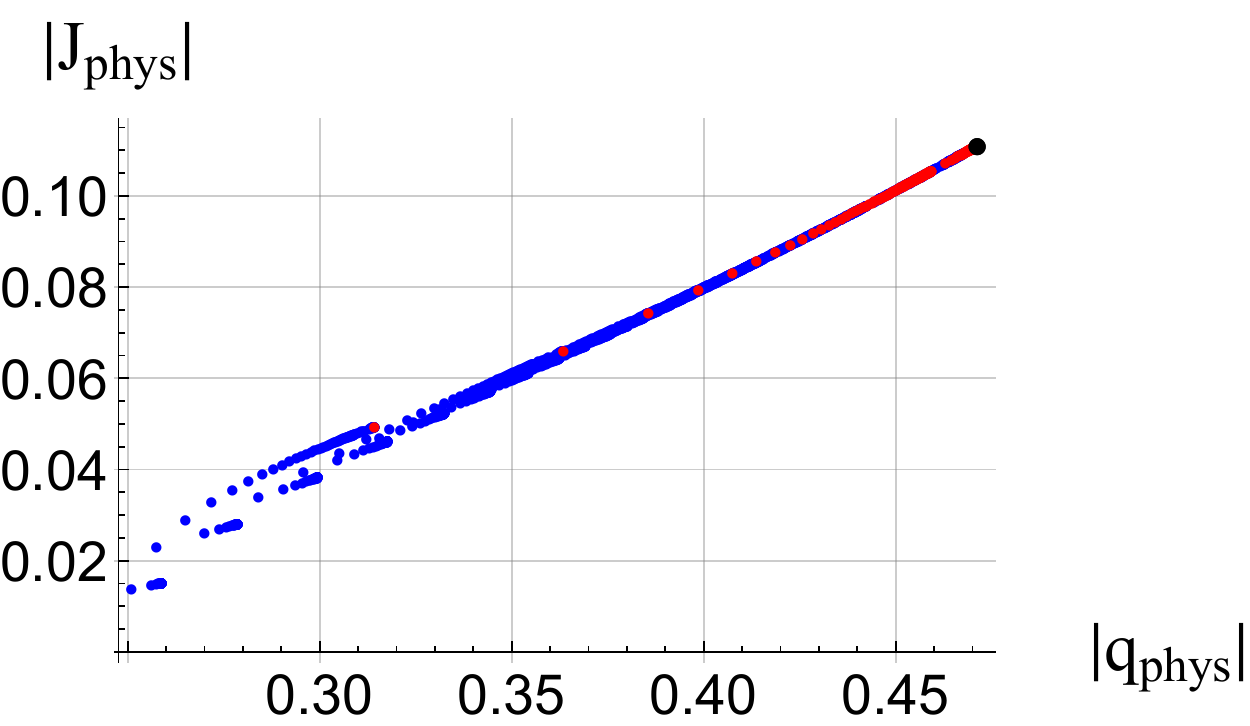}}{\includegraphics[scale=.5]{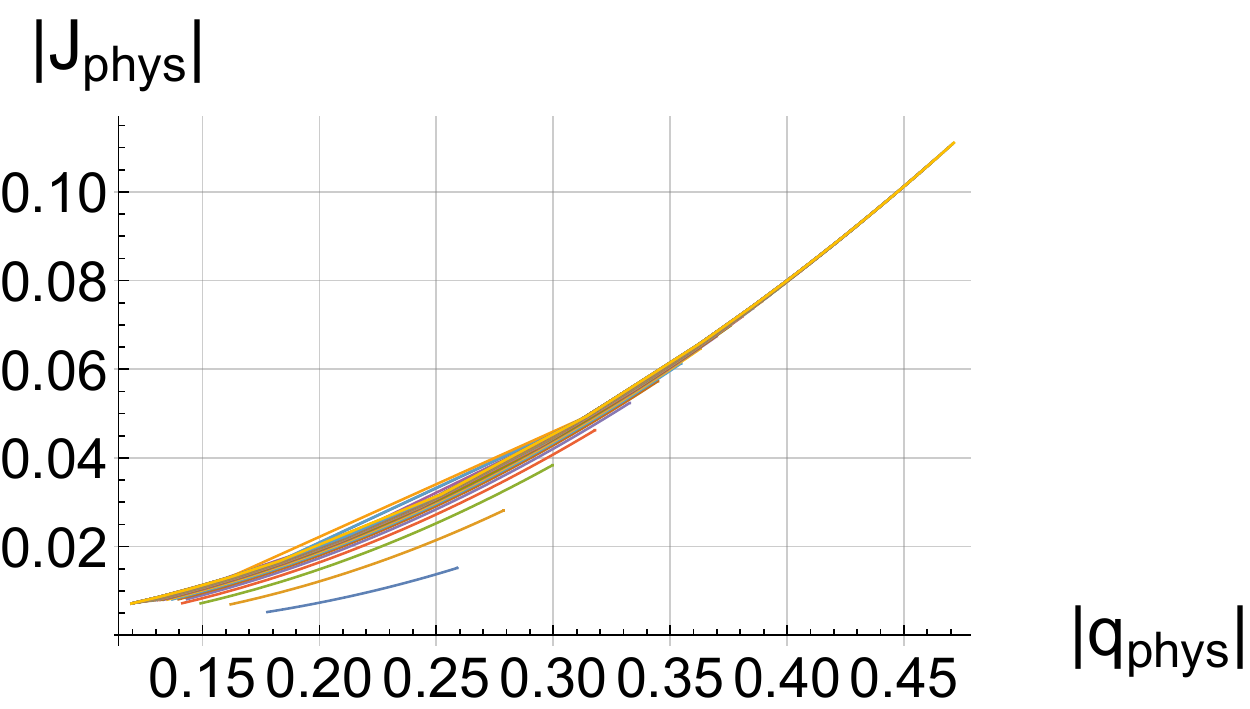}}
\caption{A comparison of the correlations between 
$\ephysphys$ and $M_\phys$  (upper plots) and $\ephysphys$ and $|J_\phys|$ (lower plots), all values scaled to $\Lambda=3$, of the solutions with integer pairs $(\newn_1,\newn_2)$ found in~\cite{ChHoerzinger} (left plots) and the current work (right plots). Left plots: The blue dots correspond to about 2000 solutions which are obtained by taking all values of $1\le \newn_2< \newn_1\le 50$ and a sample of values in the range $1\le \newn_2< \newn_1\le 1000$. The red dots are obtained by letting $\newn_1\to\infty$, with $1\le \newn_2 \le 9900$. The black dot is the limit $\newn_1\to\infty$, $\newn_2\to \infty$. Right plots: Along the plotted lines the ``quantum number'' $n=\newn_1-\newn_2$ is constant, where $\frac{1}{20} \le \newn_2< \newn_1\le 10^7$ as described above.  In the $\ephysphys$-$M_\phys$  plots $n$ is increasing from left to right, on the remaining plots from lower to higher curves. }
\label{F21VI15.1}
\end{figure}

Two plots displaying correlations between the physical parameters are shown in
Figure~\ref{F21VI15.1}.
One observes that all resulting parameters $(a,M,\ephysphys)$ are bounded, and that  the values of the parameters approach linear correlations as both $\newn_1$ and $\newn_2$ tend to infinity. The exact affine relations between the quantities of interest when $1 \ll \newn_1 \ll \newn_2$ proved in~\cite{ChHoerzinger} continue to hold in the current setting. For example, we have
\beal{21VI15.41}
 |J_\phys|
  & \approx & -\sqrt 5
  M_\phys + \frac{2}{3}
   \,,
\\
 \enosquare
  & \approx & -\frac{9}{3} M_\phys +\frac{\sqrt{2} -\sqrt{5}}{3}
   \,,
\\
 S_G
  & \approx & -\frac{\sqrt{5} \pi}{4}
  M_\phys -\frac{13}{36} \pi
   \,,
\eeal{21VI15.42}
where $S_G= -\Lambda V/(8 \pi)$ (with $V$ - the volume) is the gravitational contribution to the total action $S$ (compare~\cite[Equation~(B.8)]{ChHoerzinger}) of the Riemannian solution.

The physical parameters of the Lorentzian partner solutions are all bounded from above, cf.~Table
\ref{T8IX15.1x}. One can further check that  $|q_{phys}|$, $M_{phys}$, and $ |J_{phys}|$ tend to $0$ and $S_G$ tends to $-\pi$ when $\newn_2$ tends to zero and $\newn_1$ tends to infinity.

 \begin{table}
\caption{\label{T8IX15.1x} Maximal values of $\enosquare{}$, $M_{\phys}$, $|J_{\phys}|$, and the total action $S$ with the corresponding pairs  $(\newn_{1 },\newn_{2 })$. All values scaled to $\Lambda=3$.}
 \begin{ruledtabular}
\begin{tabular}{ | l | c | l | }
   & $(\newn_{1 },\newn_{2 })_{\max{}} $ & max.
\\ \hline
 $|q_{\phys}|$ & $(\infty,\infty)$ & $\frac{\sqrt{2}}{3} \approx 0.47$
\\
  $M_{\phys}$ & $(\infty,\infty)$ & 0.2548
\\
   $|J_{\phys}|$ & $(\infty,\infty)$ & $\frac{1}{9} \approx 0.111$
\\
  $S$ & $(\infty,\newn_2)$ & $ \infty $
\end{tabular}
 \end{ruledtabular}
 \end{table}

In particular the physical mass of the Lorentzian
           partners is strictly positive, bounded away from zero, and bounded from above.

Solutions with very large values of $\newn_1$ are strongly suppressed  when path-integral  arguments are invoked.

\def\polhk#1{\setbox0=\hbox{#1}{\ooalign{\hidewidth
  \lower1.5ex\hbox{`}\hidewidth\crcr\unhbox0}}} \def\cprime{$'$}
  \def\cprime{$'$} \def\polhk#1{\setbox0=\hbox{#1}{\ooalign{\hidewidth
  \lower1.5ex\hbox{`}\hidewidth\crcr\unhbox0}}} \def\cprime{$'$}
  \def\cprime{$'$}

\end{document}